\documentclass{mn2e}
\input{epsf}

\newcommand{\be}{\begin{equation}}
\newcommand{\ee}{\end{equation}}
\newcommand{\rhocrit}{\rho_{\rm c}} 
\newcommand{\sigate}{ \sigma_{8}}
\newcommand{\hinv}{\hbox{$h^{-1}$}} 
\newcommand{\msol}{\hbox{${\rm M}_\odot$}} 
\newcommand{\LCDM}{$\Lambda$CDM }
\newcommand{\kmsmpc}{\, \rm{km}\,  \rm{s}^{-1}\, \rm{Mpc}^{-1}}
\newcommand{\lsim}{\lower.5ex\hbox{\ltsima}}
\newcommand{\lta}{\lower.5ex\hbox{\ltsima}}
\newcommand{\ltsima}{$\; \buildrel < \over \sim \;$}
\newcommand{\gta}{\lower.5ex\hbox{\gtsima}}
\newcommand{\gtsima}{$\; \buildrel > \over \sim \;$}
\newcommand{\mpc}{\textrm{Mpc}}

\newcommand{\mtwo}{M_{200}}
\newcommand{\mback}{M_{180b}}
\newcommand{\rback}{r_{180b}}
\newcommand{\rtwo}{r_{200}}
\newcommand{\vtwo}{v_{200}}
\newcommand{\mvir}{M_{\rm vir}}
\newcommand{\rvir}{r_{\rm vir}}
\newcommand{\rhs}{r_{\rm hs}}
\newcommand{\rta}{r_{\rm ta}}
\newcommand{\rhalo}{r_{\rm halo}}
\newcommand{\rhalohat}{\hat{r}_{\rm halo}}
\newcommand{\mhs}{M_{\rm hs}}
\newcommand{\mta}{M_{\rm ta}}
\newcommand{\mhalo}{M_{\rm halo}} 
\newcommand{\aeq}{a_{\rm eq}} 
\def \kmsmpc    {{\rm\ km\ s^{-1}\ Mpc^{-1}}}

\def \etal      {\hbox{et al.} }
\def \se        {\!=\!}
\def \sims      {\sim \!}
\def \ssim      {\! \sim \!}

\def \sequiv    {\! \equiv \!}
\def \spropto   {\! \propto \!}

\title{The Ultimate Halo Mass in a \LCDM Universe}

\author[M. T. Busha, A. E. Evrard, F. C. Adams and R. H. Wechsler]
        {Michael T. Busha,$^{1}$\thanks{E-mail: mbusha@umich.edu}
        August E. Evrard,$^{1,2}$ 
        Fred C. Adams,$^{1,2}$ and
        Risa H. Wechsler$^{3, 4}$
        \\
        $^1$Michigan Center for Theoretical Physics, Department of
        Physics, 500 E. University, University of Michigan, Ann Arbor, MI,
        48109\\
        $^2$Astronomy Department, University of Michigan, Ann Arbor,
        MI 48109\\
        $^3$Department of Astronomy and Astrophysics, Kavli Institute for
        Cosmological Physics, The University of Chicago, Chicago, IL 60637\\
       $^4$Hubble Fellow, Fermi Fellow}
\begin{document}

\maketitle

\begin{abstract}

In the far future of an accelerating \LCDM cosmology, the cosmic web
of large-scale structure consists of a set of increasingly isolated
halos in dynamical equilibrium. We examine the approach of collisionless 
dark matter to hydrostatic equilibrium using a large N-body simulation
evolved to scale factor $a = 100$, well beyond the vacuum--matter
equality epoch, $\aeq \se 0.75$, and $53 h^{-1}$~Gyr into the future
for a concordance model universe $(\Omega_m\se0.3, \Omega_\Lambda\se0.7$). 
The radial phase-space structure of halos --- characterized at $a \lta
\aeq$ by a pair of zero-velocity surfaces that bracket a dynamically
active accretion region --- simplifies at $a \gta 10\aeq $ when these
surfaces merge to create a single zero-velocity surface, clearly defining
the halo outer boundary, $\rhalo$, and its enclosed mass, $\mhalo$.  This
boundary approaches a fixed physical size encompassing a mean interior
density $\sims 5$ times the critical density, similar to the
turnaround value in a classical Einstein-deSitter model.   We relate
$\mhalo$ to other scales currently used to define halo mass ($\mtwo$,
$\mvir$, $\mback$) and find that $\mtwo$ is approximately half of the
total asymptotic cluster mass, while $\mback$ follows the evolution of
the inner zero velocity surface for $a \lta 2$ but becomes much larger
than the total bound mass for $a \gta 3$.  The radial density profile
of all bound halo material is well fit by a truncated Hernquist
profile.  An NFW profile provides a somewhat better fit interior to
$\rtwo$ but is much too shallow in the range $\rtwo < r < \rhalo$. 

\end{abstract}

\begin{keywords}
cosmology: theory --- large scale structure of the universe 
--- dark matter 
--- dark energy
\end{keywords}

\section{Introduction} 

In cosmologies in which density fluctuations are seeded by an early
inflationary period (e.g., Kolb \& Turner 1990), gravity acts during
the era of cold dark matter domination to evolve a ``cosmic web'' of
non-linear structure (e.g., Bond, Kofman, \& Pogosyian 1996) that can
be approximately described as a set of roughly spherical halos, each
characterized by a mass $M$.  Within a given cosmology, the spatial
density, clustering properties, and internal structure of such halos
evolve with time in ways that are becoming increasingly well understood
(e.g., Cooray \& Sheth 2002, Kravtsov et al 2004).

Oddly, the central element of this picture, the halo mass, remains
poorly defined.  During the growth phase of the web, the mass
distribution of a halo smoothly connects to the cosmological
background of adjoining filaments, sheets, and voids.  Any given halo
is a mix of `old' halo material that may be near hydrostatic and
virial equilibrium and `new' material gained from recent
accretion. Although a radial gradient in the ratio of these two
components is present, no clear edge that would allow a unique
definition of mass separates them.  In contrast, analytic collapse
models based on spherical symmetry possess a well-defined outer shock
or caustic surface (Bertschinger 1985; Filmore \& Goldreich 1984) that
emerges at a characteristic density.  These models have motivated
several mass definitions based on the enclosed density.  White (2001)
offers a recent discussion of these and other mass definitions applied
to large-scale structure simulations.

In contrast to the present epoch, the far future of
a \LCDM universe provides an opportunity to cleanly define halo mass.
In the relatively near cosmological future, merger activity comes to
an effective end and dark matter halos evolve toward a dynamically
quiet state (Busha \etal 2003, hereafter paper I; Nagamine \& Loeb
2003; see also the reviews of Adams \& Laughlin 1997; Cirkovic 2003).
In this paper, we show that the phase-space configuration of dark
matter halos reaches a well-behaved asymptotic state characterized by
a single zero-velocity surface that uniquely defines the halo edge.
Essentially all material internal to this surface is bound to and
equilibrated within the halo, whereas material outside this surface is
expanding away with the locally perturbed Hubble flow.  In \S2, we
describe the simulation used and the mass measures employed in our
analysis.  Results are given in \S3, followed by brief summary and
discussion section (\S4). Our mass and radius measurements assume a
Hubble constant  $H_0 = 100 h \kmsmpc$ with $h = 0.7$. 

\section{Simulations and Mass Measures} 

We present results based on a simulation run with L-Gadget, a
specialized version of the N-body code GADGET (Springel, Yoshida, \&
White 2001).  L-Gadget simulates only
collisionless matter in a manner that optimizes cpu and memory
resources.  Our simulation models a patch of flat space in a \LCDM
universe with current matter density $\Omega_{m} = 0.3$, vacuum
density $\Omega_{\Lambda} = 0.7$, and power spectrum normalization
$\sigate = 0.9$, values consistent with current observational data
(Spergel \etal 2003).

The dark matter in a periodic cube of side length $200 \hinv \rm{Mpc}$
is modeled by $256^3$ particles of mass $3.97 \times 10^{10} \hinv
\msol$. The simulation was started at redshift $z = 19$ (scale factor
$a = 0.05$), evolved through the present epoch ($a = 1$), and
continued forward to $a = 100$.  A gravitational softening parameter
fixed at $40 \hinv \rm{kpc}$ (in physical units) was used throughout
the computation.  The simulation was run on 8 dual-cpu nodes of a
Beowulf cluster at the University of Michigan.  A total of 300 outputs
equally spaced in log$(a)$ were stored from the run.  As the scale
factor increases, the universe evolves from being matter dominated to
vacuum energy dominated.  We define $\aeq$ as the epoch at which the
energy densities in the two components are equal.  For the chosen
model, this transition occurs at $\aeq \se
(\Omega_m/\Omega_{\Lambda})^{1/3} \se 0.75$. 

We identify dark matter halos using a standard percolation (or
friends-of-friends, FOF) algorithm with linking length 0.15 times the
inter-particle spacing. The halo center is identified as the most
bound particle of the resulting group, and the halo velocity is
defined as the center of mass velocity of the linked set of
particles. At the end of the simulation this algorithm identified
about 2900 halos with at least 500 particles.  The largest halo at
that time contains 83,600 particles, equivalent to $3.3\times 10^{15}
\hinv \msol$.  In the analysis below, we measure ensemble properties
using the 400 most massive halos.

The FOF algorithm defines halo centers about which we measure several
enclosed mass scales. The first mass scale is $\mtwo$, the mass
contained within a sphere of radius $\rtwo$ that encloses a mean density
200 times the critical density $\rhocrit(a)$ at the epoch of interest.
In addition, a so-called virial mass $\mvir$ (within $\rvir$) is
defined to enclose a mean density $\Delta_c(a)$ times the critical
density, where $\Delta_c(a)$ is an epoch-dependent threshold based on
a simple, spherical collapse model (e.g., Eke, Cole, \& Frenk 1996).
$\Delta_c$ reaches a constant value of $20.4$ as
$\Omega_m(a) \rightarrow 0$.  A third mass scale $\mback$ is
defined by a mean enclosed density of 180 times the {\it background}
mass density $\rho_m(a)$.  

\section{Results} 

\begin{figure}
\begin{center}
    \leavevmode
    \epsfxsize=8cm
    \epsfbox{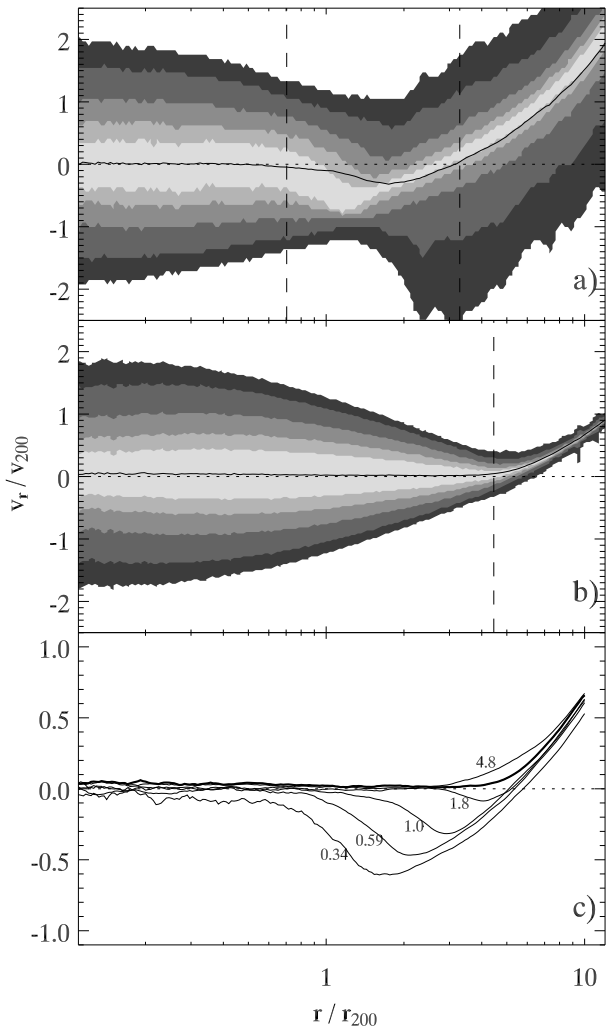}
\end{center}
\caption{The distribution of dark matter radial velocities as a
function of distance from the halo center.  The top two panels
show the conditional phase-space density $p(v_r|r)$ as a function of
radius for the ensemble of 400 largest halos at the present epoch (a)
and for the future when $a = 100$ (b). The solid line shows the mean
velocity as a function of radius; the grey scale indicates the regions
enclosing 40, 60, 80, 95, and 99 percent of the particle population
as specified by $p(v_r|r)$; the vertical lines represent the
zero-velocity surfaces.  Panel (c) shows the mean radial velocity for
an ensemble of halos at epochs $a$ = 0.34, 0.59, 1.0, 1.8, and 4.8,
with the bold line representing the function at $a = 100$.} 
\label{fig:phasefig}
\end{figure}

Previous studies (paper I; Nagamine \& Loeb 2003) have shown
that large-scale structure quickly approaches a stable configuration
in the future deSitter phase of a \LCDM cosmology.  Mergers and
accretion slow dramatically after the scale factor exceeds $a = 2-3$,
signifying the end of the dynamically active stage of the cosmic web.
The change in dynamical activity, from active to absent, is apparent
in the reduced phase-space density $f(r,v_r)$ of halos as shown in
Figure~\ref{fig:phasefig}. This figure plots the proper radial
velocity $v_r$ against distance from the halo center $r$ for an
ensemble average of the 400 most massive halos using $\rtwo$ and
$\vtwo = \sqrt{G \mtwo/\rtwo}$ as scale variables. 

Figure~\ref{fig:phasefig}a shows the conditional probability
$p(v_r\,|\, r) = f(r,v_r)/\rho(r)$ for these 400 halos at $a = 1$.
The solid line gives the mean radial velocity, while the grey scale
regions delimit velocities containing $40\%$, $60\%$, $80\%$, $95\%$,
and $99\%$ of the material at each radius. This ensemble average phase
space density can be divided into three principal regions: i) an inner
hydrostatic core ($\langle v_r \rangle = 0$) that is relatively well
relaxed; ii) an intermediate accretion envelope ($\langle v_r \rangle
< 0$) containing two opposing streams of material, one on its first
inward journey toward the halo center and the other passing outward
after pericentric passage; and iii) an outflow region ($\langle v_r \rangle >
0$) dominated by the locally perturbed Hubble flow.  Nearly all of the
material in regions i) and ii) is gravitationally bound to the halo (a
modest fraction of material within $\rtwo$ can be scattered out of a
halo, see Fig.~3 of paper I) while essentially all of region iii) is
unbound. The characteristic scale that separates the outflow and
accretion regions is often called the turnaround radius $\rta$ (Gunn
\& Gott 1972), a term motivated by spherical models of expanding mass
shells.  We call the radius that separates the two interior zones the
hydrostatic radius $\rhs$ and measure its value using a threshold
condition for the binned, mean radial velocity.  Starting from the
interior, we identify $\rhs$ as the minimum radius at which  $|\langle
v_r \rangle/\vtwo| > 0.1$.  The turnaround radius is identified in the
same manner, identifying the maximum radius at which $|\langle v_r
\rangle/\vtwo| < 0.1$. At $a=1$, the values of these characteristic
radii are $\rhs = 0.70 \rtwo$ and $\rta = 3.3 \rtwo$ and are marked by
the vertical dashed lines in Figure~\ref{fig:phasefig}a. 

The averaging process used to create the density in
Figure~\ref{fig:phasefig}a blends the effects of substructures
within individual halos, especially in the hydrostatic region.  At a
given radius, this region has a nearly Gaussian distribution of radial
velocities with zero mean, signatures of hydrostatic and virial
equilibrium.  Due to the presence of pairs of massive clusters, the
outflow region for the ensemble profile has a rather broad dispersion.
The fact that the distribution of radial velocities for the outflow
region is more sharply peaked than in the virialized region indicates
that the surrounding regions generally contain halos that 
are less massive than those of our high-mass selected sample. 

The situation in the far future is markedly different from the
present.  Figure~\ref{fig:phasefig}b shows the ensemble phase-space
density at $a = 100$.  The most striking changes are the disappearance
of the accretion region and the dramatic cooling of the outflow
region. Here, $\rhs$ and $\rta$ have merged to form a single
zero-velocity surface at $\rhalo = 4.5 \rtwo$, represented by the
dashed vertical line.  In addition, relatively few nearby halos are
present to disrupt the outflow stream.  Neighboring halos that existed
at $a = 1$ have either merged or been pulled away in the deSitter
expansion. Although not shown here, the phase-space density of the
single most massive halo at $a = 100$ is nearly identical to the
ensemble averaged profile. Because mergers cease long before $a =
100$, the substructure present within any individual halo has several
dynamical times to relax via tidal stripping, phase mixing, and
dynamical friction (see paper I for a comparison with the phase-space
distributions for individual halos).

The elimination of the accretion region at $a =100$ is accompanied by
an expansion of the hydrostatic region. Figure~\ref{fig:phasefig}c
shows the evolution of the average $v_r / \vtwo$. The bold curve shows
the  final profile at $a = 100$; the thinner curves depict the epochs
$a$ = 0.34, 0.59, 1.0, 1.8, and 4.8. The extent of the accretion
region decreases with time as $\rhs/\rtwo$ grows and $\rta/\rtwo$
slightly shrinks.  Note also that there is an expulsion epoch at $a =
4.8$, shortly after mergers have ended.  At this time, $v_r / \vtwo$
is somewhat larger than its asymptotic value in the range $r / \rtwo =
3 - 8$ because the unbound particles from the final mergers are being
ejected.

\begin{figure}
\begin{center}
    \leavevmode
    \epsfxsize=8cm
    \epsfbox{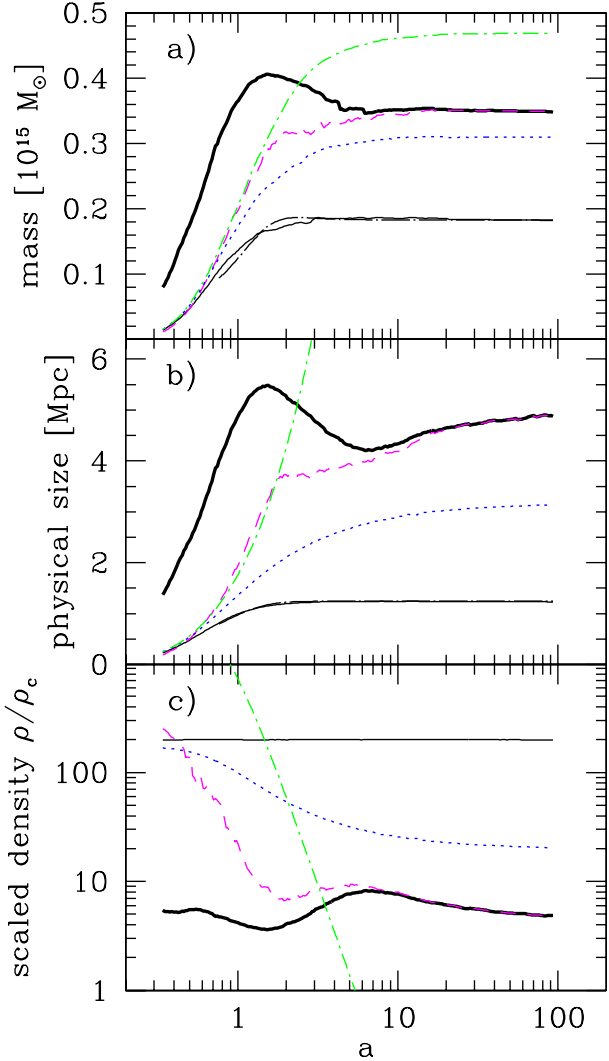}
\end{center}
\caption{Comparison of mass and radial scales for the progenitors
of the 400 most massive halos identified at the end of the computation.
The four panels show the scale factor dependence of (a) mean masses; (b)
mean physical sizes; (c) enclosed densities relative to the critical value.
The line styles indicate the different mass measures: $\mta$ (bold);
$\mhs$ (dashed); $\mtwo$ (solid); $\mvir$ (dotted); and $\mback$
(dot-dashed). The line styles for the radial scales are analogous. In
panel (b), the dot-long-dashed curves show the asymptotic form of
$\rtwo$ (see text) for $a>\aeq\se 0.75$. } 
\label{fig:mta}
\end{figure}

The time evolution of this transition is presented in
Figure~\ref{fig:mta}.  As before, we show ensemble behavior of the 400
most massive halos selected at $a \se 100$, with halos at earlier
epochs restricted to the most massive progenitors of this final
population (where most massive progenitor is defined to be the halo
with the largest $M_{FOF}$ that contributes at least 20\% of its
particles to its descendant).  For each halo and epoch, we identify
the physical values of the characteristic radii, ($\rhs$, $\rta$,
$\rtwo$, $\rvir$, $\rback$) along with the respective enclosed masses.
To determine $\rhs$ and $\rta$ for each halo, we first time smooth the
individual profile by co-adding the halo configuration over seven
consecutive outputs.  We then measure $\rta$ using a linear
extrapolation of the outflow over a factor of six in $v_r$, starting
at the radius $r = 10 \mpc$ and working inward.  Applying a threshold (as
above) instead of extrapolating produces somewhat larger values for
$\rta$, but causes only a small change in the values for $\mta$.  We
choose the extrapolation method because it provides less noisy
estimates during the early phase of active halo growth.  The
hydrostatic region, $\rhs$ is measured using a threshold technique on
the mean radial velocity measured in radial bins, identical to the
method used for Figure~\ref{fig:phasefig}.  We use logarithmically
spaced bins, 30 per decade, and identify $\rhs$ as the radius at which
$|v_r|/ v_{ta} > 0.1 {\rm,} \ v_{ta} = \sqrt{G \mta / \rta}$.
Typically, $v_{ta} \sim 0.7 \vtwo$.

Figure~\ref{fig:mta} shows the time evolution of the mean physical
radii and enclosed masses for the different measures.  During the
early Einstein-deSitter phase, $a < \aeq$, the sizes defined by mean
interior densities $\rtwo$, $\rvir$ and $\rback$ are similar, whereas
the hydrostatic boundary $\rhs$ lies somewhat interior to $\rtwo$ and $\rvir$. 
The turnaround radius $\rta$ lies well beyond these scales while the
growth of linear perturbations remains robust.  

The interior scales diverge at epochs $a > \aeq$ when vacuum effects
dominate.  As linear growth stagnates, large-scale mass density
fluctuations $\delta\rho/\bar{\rho}$ become frozen into the
background.  Continuity of the density field implies that the mean
size $\rback$ (defined relative to the mean matter density) becomes
constant in the comoving frame, so the physical size grows
exponentially in time. In contrast, the ensemble mean $\rtwo$ rapidly
approaches its asymptotic value, following $\rtwo(a) \ = \ r_{{\rm
200},\infty} \  [1-{\rm exp}(-(a/\aeq)^{1.5})] \,$ at late times, with
$r_{{\rm 200},\infty} \se 1.2$~Mpc for this high-mass sample.  This
fit is shown for $a>\aeq$ by the dot-long-dashed line in
Figure~\ref{fig:mta}b. 

The value of $\rvir$ relaxes less quickly, due to the decline in the
variable threshold $\Delta_c(a)$ as the scale factor increases.  The
enclosed masses $\mtwo$, $\mvir$ and $\mback$ grow monotonically until
reaching $99\%$ of their asymptotic limits at $a \se 2.9$, $8.6$ and
$13.6$, respectively, corresponding to ages of $32$, $52$ and $60$
Gyr. The innermost mass scale $\mtwo$ experiences a slight decline at
late epochs, $\spropto (0.00045 \pm 0.00015) {\rm ln}(a)$, but it is
not clear whether this drift is physical or numerical.

The hydrostatic boundary $\rhs$ tracks $\rback$ until $a \sims 2$, then
slows its growth and relaxes toward an asymptotic value.  
The turnaround radius also grows rapidly until $a \simeq 1.5$, then
declines until $a \sims 5$, and increases slowly thereafter.  The
decline in size and enclosed mass from the peak until $a \ssim 5$
arises from a change from infall to outflow in the accretion regime
lying between $\rhs$ and $\rta$, as indicated by the outflow
enhancement at this epoch in Figure~\ref{fig:phasefig}c.  At $a
\ssim 1$, accretion just outside the hydrostatic boundary drops
dramatically.  After a crossing time, the modest mass fraction of this
accreted material that is scattered to unbound orbits emerges at radii
$r > \rhs$ with positive radial velocity at $a \sims 2-5$.  This
material escapes from the halos, but we have not investigated its
ultimate fate. Presumably, some fraction emerges as the cores of
stripped sub-halos that remain bound, whereas the remainder may emerge
sufficiently tenuous and hot that it never collapses into a halo of
cosmologically interesting mass.  Future studies using higher
resolution experiments are needed to address this question.

For test particles in Hubble flow around a halo of mass $M$, the outer
zero-velocity surface location can be estimated using a Newtonian
binding energy, with the result (eq.~[11] of paper I) $r \simeq
(M/10^{12} \msol)^{1/3} \mpc$.   Applying this estimate using the mean
asymptotic halo mass in Figure~\ref{fig:mta}a yields a value of $5
\mpc$, close to the asymptotic mean size shown in Figure~\ref{fig:mta}b.
We caution that the values of $\rhs$ and $\rta$ are sensitive at the
$\sims 10\%$ level to the choice of threshold and/or the
interpolation scheme used to locate the zero-crossing.  However, the
steep behavior of the density profile near the boundary ($\rho
\spropto r^{-\gamma},\ \gamma \gg 4$) leads to a much smaller 
uncertainty ($\lta 2\%$) in enclosed mass. 

The merging of the turnaround and hydrostatic scales signals the end
of high mass structure formation, and therefore the end of the growth
phase of the cosmic web.  Thereafter, halos maintain a fixed physical
size and the morphology of the cosmic web, when viewed in the comoving
frame at a density threshold near critical, becomes less of a web and
more an increasingly fine spray of droplets (see paper I and Nagamine
\& Loeb 2003).  From Figure~\ref{fig:mta}a, the ratio $\mhs / \mta$
reaches $99\%$ of its asymptotic value at $a \se 7.4$ ($a/\aeq \se
10$).  At epochs beyond $10 \aeq$, we can clearly define the
(ultimate) halo mass as $\mhalo \sequiv \mhs \se \mta$, i.e., {\sl the
mass enclosed within the single  zero-velocity surface is the ultimate
halo mass. }

Figure~\ref{fig:mta}c shows that the turnaround/ultimate halo mass
is defined by a density threshold $\rho_{\rm halo}$ relative to
critical that lies at all times within a factor of two range spanning
$\sims (5-10) \rhocrit(a)$.  The threshold at early times lies close
to the canonical $9\pi^2/16\se5.5$ value expected from spherical
collapse in an Einstein-deSitter universe (Peebles 1980).  After
$\aeq$, the ratio $\rho_{\rm halo}/\rhocrit(a)$ first drops while the
mean halo mass peaks, then climbs to $~10$ during the period $a/\aeq
\se 2-10$ when the mean halo mass drops.  At late times, $\rho_{\rm
halo}$ is declining weakly, and its asymptotic limit, although
not well determined by this simulation, is only about $12\%$ shy of the
canonical value.  

While the gap between $\rback$ and $\rhalo$ expands exponentially in
the far future, the enclosed mass ratio converges to $\mback/\mhalo
\approx 1.35$.  While thresholding with respect to the background mass
density extends beyond the halo edge at late times, the scales defined
relative to the critical density pick out mass shells interior to the
halo: $\mvir \se 0.89 \mhalo$ and $\mtwo \se 0.52 \mhalo$.  These
particular values are sensitive to the asymptotic form of the radial
density distribution. 

\begin{figure}
\begin{center}
    \leavevmode
    \epsfxsize=7.5cm
    \epsfbox{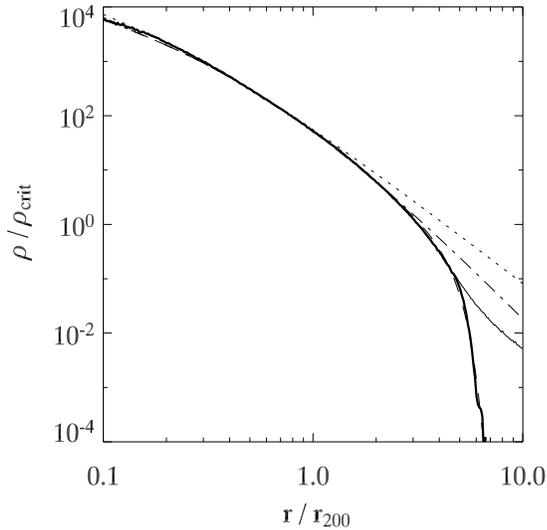}
\end{center}
\caption{The asymptotic form of the density distribution for dark 
matter halos. The solid curves show the density profile for an ensemble
average of the 400 most massive halos in the simulation at scale
factor $a$ = 100. The upper solid curve shows the total density  as a
function of radius, whereas the lower solid curve includes only  the
bound particles. The dotted curve shows the best fit NFW profile.  The
dot-dashed curve shows the best fit Hernquist profile, and the  dashed
curve represents a truncated Hernquist profile (see
eq. [\ref{eq:rhohalo}]). } 
\label{fig:dprofile}
\end{figure}

Figure~\ref{fig:dprofile} shows the mean radial profile obtained at $a
\se 100$ from the stacked ensemble of the 400 most massive halos.  We
use $\rtwo$ as the scale radius, but results are similar when other
characteristic scales are used.  The light solid line shows the mean
profile while the heavy solid line shows the profile for material
bound to each halo, using specific energy $E/m \se v^2/2 - GM(<r)/r +
(4\pi/3)G \rho_{\Lambda} r^2$ (with $v$ and $r$ the proper velocity
and radius, respectively).  The bound material is well fit over the
entire halo volume by a truncated Hernquist (1990) profile 
\be 
\rho(r) \ = \ {\rho_0 \over (r/r_c) (1 + r/r_c)^3} \ 
e^{-(r/\rhalohat)^{5.6}}  \, , 
\label{eq:rhohalo}
\ee 
with characteristic radius $r_c \se 0.62 \rtwo$ and truncation
scale $\rhalohat \se 4.6 \rtwo$.  The latter measure of halo size
agrees extremely well with the value of $4.5 \rtwo$ obtained from the
ensemble average zero-velocity surface.  This fit has a least-square
error of about 12\%.  For comparison, an NFW profile $\rho(r) = 4
\rho_s (r/r_s)^{-1} [1 + (r/r_s)]^{-2}$ (Navarro, Frenk, \& White
1997), obtained by fitting within $\rtwo$ is shown in
Figure~\ref{fig:dprofile} by the dotted line.  The profile with $r_s
\se 0.25 \rtwo$ fits the inner regions of the halo well, but it is
much too shallow beyond $\rtwo$, overestimating the halo density at
$\rhalo$ by an order of magnitude.  Note that simple scaling for the
evolution of the concentration proposed by Bullock et al (2001),
$c_{vir}(a) \sim a$, breaks down as the growth rate becomes constant
and the profile out to $\rvir$ becomes fixed after a few Hubble
times.

\section{Summary} 

From a large N-body simulation that follows the long term evolution of
collisionless structure in a \LCDM cosmology, we have examined the
ultimate approach to equilibrium for a sample of 400 massive halos
with final mean mass $3.5 \times 10^{14} \msol$. During the
matter-dominated era ($a \lta \aeq \se 0.75$), the radial phase-space
structure of the halos is complex, consisting of an inner hydrostatic
region, an intermediate accretion zone, and an exterior region that
expands with the perturbed Hubble flow. During the interval $\aeq \le a \le
10 \aeq$, accretion shuts down and the intermediate region is briefly
dominated by outflow rather than infall.  For $a \gta 10\aeq$, the
intermediate region disappears, and the hydrostatic and turnaround 
scales merge to form a single zero-velocity surface that provides an
unambiguous definition of halo mass.

The existence of multiple mass scales commonly used in the literature
to describe clusters is a direct reflection of the complexity of the
accretion region during the growth phase of structure. This work 
illuminates the evolving relationships between the hydrostatic and
turnaround scales of halos and scales defined by mean interior density
thresholds.  Although thresholding with respect to the background
density ($\mback$) has advantages for calculating the halo space
density (Jenkins \etal 2001), its use to describe future structure is
compromised by our finding that $\mback$ exceeds the asymptotic halo
mass for $a \gta 2$.  Masses defined by thresholds relative to the
critical density converge to well-defined values.  For $a \gta 10 \,
\aeq$, we find $\mhalo \se 1.1 \mvir \se 1.9 \mtwo$.  

Using only bound material, the
ensemble mean density profile of the 400 most massive halos is well
fit by a modified Hernquist model, with scale radius $0.62\rtwo$ and
truncation radius $4.6\rtwo$ (eq.~[\ref{eq:rhohalo}]).  The origin of
this particular form, as well as other issues such as its dependence
on halo mass and its extension to an ellipsoidal description, remains
to be investigated. 

\bigskip 

It is a pleasure to thank Volker Springel for the use of L-Gadget as
well as Tony Bloch for useful conversations. This work was supported
by the Michigan Center for Theoretical Physics, by NASA through
grants NAG5-13378 and NNG04GK56G0, and by the National Science
Foundation under Grant No. PHY99-0794 to the Kavli Institute for
Theoretical Physics.

\end{document}